\documentclass[a4paper,twocolumn,accepted=2024-03-13]{quantumarticle}
\pdfoutput=1

\usepackage[utf8]{inputenc}
\usepackage[bbgreekl]{mathbbol}
\usepackage{amsfonts}
\DeclareSymbolFontAlphabet{\mathbb}{AMSb}
\DeclareSymbolFontAlphabet{\mathbbl}{bbold}

\usepackage{gensymb,amssymb,graphicx,color,amsmath,mathtools,bm}
\usepackage{hyperref}
\usepackage[normalem]{ulem}
\usepackage[caption=false]{subfig} % position=top
\usepackage{braket}

\newcommand{\secref}[1]{Sec.\,\ref{#1}}
\newcommand{\refcite}[1]{Ref.\,\cite{#1}}

\newcommand{\eqnref}[1]{Eq.\,\eqref{#1}}
\newcommand{\eqsref}[1]{Eqs.\,\eqref{#1}}
\newcommand{\figref}[1]{Fig.\,\ref{#1}}
\newcommand{\figsref}[1]{Figs.\,\ref{#1}}
\newcommand{\sfigref}[2]{Fig.\,\hyperref[#1]{\ref{#1}#2}}

\definecolor{kspink}{RGB}{200,0,200}

\newcommand{\bKet}[1]{\big|#1\big\rangle}
\newcommand{\bBraket}[1]{\big\langle #1 \big\rangle}

\hypersetup{
  colorlinks = true,
  urlcolor = blue,
  pdfauthor = {Kevin Slagle},
  pdftitle = {Slagle - 2023 - The Gauge Picture of Quantum Dynamics}
}

\begin{document}

\title{The Gauge Picture of Quantum Dynamics}

\author{Kevin Slagle}
\affiliation{Department of Electrical and Computer Engineering, Rice University, Houston, Texas 77005 USA}
\affiliation{Department of Physics, California Institute of Technology, Pasadena, California 91125, USA}
\affiliation{Institute for Quantum Information and Matter and Walter Burke Institute for Theoretical Physics, California Institute of Technology, Pasadena, California 91125, USA}

\begin{abstract}
Although local Hamiltonians exhibit local time dynamics,
  this locality is not explicit in the Schr\"{o}dinger picture in the sense that the wavefunction amplitudes
  do not obey a local equation of motion.
We show that geometric locality can be achieved explicitly in the equations of motion
  by ``gauging'' the global unitary invariance of quantum mechanics into a local gauge invariance.
That is, expectation values $\braket{\psi|A|\psi}$ are invariant under a global unitary transformation acting on
  the wavefunction $\ket{\psi} \to U \ket{\psi}$ and operators $A \to U A U^\dagger$,
  and we show that it is possible to gauge this global invariance into a local gauge invariance.
To do this, we replace the wavefunction with a collection of \emph{local wavefunctions} $\ket{\psi_J}$,
  one for each patch of space $J$.
The collection of spatial patches is chosen to cover the space;
  e.g. we could choose the patches to be single qubits or nearest-neighbor sites on a lattice.
Local wavefunctions associated with neighboring pairs of spatial patches $I$ and $J$ are
  related to each other by dynamical unitary transformations $U_{IJ}$.
The local wavefunctions are local in the sense that their dynamics are local.
That is, the equations of motion for the local wavefunctions $\ket{\psi_J}$ and connections $U_{IJ}$
  are explicitly local in space and only depend on nearby Hamiltonian terms.
(The local wavefunctions are many-body wavefunctions and have the same Hilbert space dimension as the usual wavefunction.)
We call this picture of quantum dynamics the \emph{gauge picture} since it exhibits a local gauge invariance.
The local dynamics of a single spatial patch is related to the interaction picture,
  where the interaction Hamiltonian consists of only nearby Hamiltonian terms.
We can also generalize the explicit locality to include locality in local charge and energy densities.
\end{abstract}

\newpage

\maketitle
{
  \hypersetup{linkcolor=black}
  \tableofcontents
}
\newpage
\section{Introduction}

Locality is of fundamental importance in theoretical physics.
The observable physics of quantum dynamics is local if the Hamiltonian is geometrically local\footnote{%
    Geometric locality is not to be confused with the weaker notion of $k$-locality,
      for which an operator is $k$-local if it acts on at most $k$ qubits \emph{anywhere} in space.},
  i.e. if the Hamiltonian is a sum of local operators.
An operator (other than the Hamiltonian)
  is said to be local if it only acts on a small region of space.
But locality is not explicit in the Schr\"{o}dinger picture because
  the wavefunction is global in the sense that it can not be associated with any local region of space,
  and the time dynamics of the wavefunction globally depends on all Hamiltonian terms.
On the other hand, locality is explicit in the Heisenberg picture,
  for which the time dynamics of local operators only depends on nearby local operators
  (when the Hamiltonian is local). \cite{DeutschHayden}
This motivates us to ask:
  \emph{Is it possible to modify the Schr\"{o}dinger picture to make locality explicit
  in the equations of motion?}

Gauge theory is another fundamental concept in theoretical physics.
Gauge theory is the foundation of the Standard Model of particle physics
  and is also used to describe exotic phases of condensed matter \cite{LevinWenStringNet, DQCP}.
An important tool that gauge theory provides is the gauging process,
  in which one promotes a global symmetry into a local gauge symmetry by coupling the original model to gauge fields.
For example, a scalar field theory $\mathcal{L} = \frac{1}{2} (\partial_\mu \phi)^2$
  is invariant under a global $U(1)$ symmetry $\phi(x) \to \phi(x) + \lambda$.
By coupling $\phi$ to a gauge field $A_\mu$ as in $\mathcal{L}_\text{gauged} = \frac{1}{2} (\partial_\mu \phi - A_\mu)^2$,
  the global symmetry is promoted to a local gauge symmetry where
  $\phi(x) \to \phi(x) + \lambda(x)$ and $A_\mu(x) \to A_\mu(x) + \partial_\mu \lambda(x)$.
Gauging more exotic symmetries leads to more exotic physics;
  e.g. gauging spatial symmetries can lead to gravity
  and gauging fractal symmetries can result in fracton topological order \cite{yoshidaFracton}.
In quantum mechanics, expectation values $\braket{\psi|A|\psi}$ are invariant under a
  global unitary transformation acting on the
  wavefunction $\ket{\psi} \to U \ket{\psi}$ and operators $A \to U A U^\dagger$.
Although this transformation is typically viewed as a global invariance rather than a global symmetry,
  we can still ask:
  \emph{Is it possible to gauge the global unitary invariance in quantum mechanics?
  And what are the consequences of doing so?}

We find that the answer to both questions is yes,
  and that one consequence of gauging the global unitary invariance
  is that locality becomes explicit in the equations of motion.
In order to achieve this, we introduce a collection of \emph{local wavefunctions} $\ket{\psi_I}$,
  each associated with a local patch of space.
Each local wavefunction is an element of the same Hilbert space as the usual wavefunction.
The local wavefunction associated with nearby patches are related by unitary transformations $U_{IJ}$,
  which are also dynamical.
Locality is explicit in the sense that the equations of motion [\eqnref{eq:localSchrodinger}]
  for the dynamical variables ($\ket{\psi_I}$ and $U_{IJ}$)
  only depends on nearby Hamiltonian terms and nearby dynamical variables.
(Note that locality is not explicit in Schr\"odinger's picture in this way.)

The Schr\"odinger and Heisenberg pictures are related by a time-dependant unitary transformation that acts on the wavefunctions and operators.
Our new picture of quantum dynamics is related to the Schr\"odinger and Heisenberg pictures via a local gauge transformation.
We describe these derivations in detail in \secref{sec:Heisenberg}
  and \secref{sec:gaugePicture}.
The new local equations of motion are given in \eqnref{eq:localSchrodinger},
  and the local gauge transformation is given in \eqnref{eq:gaugeSymmetry}.

In \secref{sec:interactionPicture},
  we show that the gauge picture local wavefunction associated with a patch
  is equivalent to the interaction picture wavefunction when the interaction Hamiltonian is the sum over Hamiltonian terms that have some support on that patch.
To gain intuition, in \secref{sec:circuits}
  we consider the example of a quantum circuit in the gauge picture.
In \secref{sec:measurements},
  we describe the measurement process in the gauge picture.
Although local unitary dynamics are explicitly local in the gauge picture,
  the affect of local measurement is not explicitly local in the gauge picture (similar to the Schr\"odinger picture).
In \secref{sec:patches}, we generalize spatial locality in the gauge picture to e.g. locality in local particle number or local energy density.

\section{Schr\"{o}dinger to Heisenberg Picture}
\label{sec:Heisenberg}

To warm up, we first derive the Heisenberg picture from the Schr\"{o}dinger picture.
In the Schr\"{o}dinger picture, the wavefunction evolves according to
\begin{equation}
  \partial_t \Ket{\psi^\text{S}(t)} = -i H^\text{S}(t) \Ket{\psi^\text{S}(t)} \label{eq:schrodinger}
\end{equation}
  where $H^\text{S}(t)$ is the Hamiltonian (and we set $\hbar=1$).
We use superscripts ``S'' and ``H'' to respectively label time-dependent variables
  in the Schr\"{o}dinger and Heisenberg pictures.

To derive the Heisenberg picture,
  consider applying a time-dependent unitary transformation to the wavefunction and operators:
\begin{equation}
\begin{aligned}
  \Ket{\psi^\text{H}(t)} &= U^\dagger(t) \Ket{\psi^\text{S}(t)} \\
  A^\text{H}(t) &= U^\dagger(t) A^\text{S}(t) U(t)
\end{aligned} \label{eq:Htransform}
\end{equation}
Equation~\eqref{eq:Htransform} defines the meaning of the ``H'' superscript for all operators and wavefunctions.
This unitary transformation has the essential property that it does not affect expectation values:
\begin{equation}
    \Braket{\psi^\text{H}(t) | A^\text{H}(t) | \psi^\text{H}(t)}
  = \Braket{\psi^\text{S}(t) | A^\text{S}(t) | \psi^\text{S}(t)}
\end{equation}
Since $U(t)$ does not affect the physics,
  the unitary transformation $U(t)$ could therefore be viewed as a global ``gauge'' transformation,
  which we will use to move the time dynamics from the wavefunction to the operators.

Let $G^\text{S}(t)$ be the Hermitian operator such that
\begin{equation}
  \partial_t U(t) = -i G^\text{S}(t) U(t) \label{eq:dUG}
\end{equation}
Or equivalently,
  $\partial_t U(t) = -i U(t) G^\text{H}(t)$
  where $G^\text{H}(t) = U^\dagger(t) G^\text{S}(t) U(t)$ is defined by \eqnref{eq:Htransform}.
The time derivatives of the new wavefunction and operators are
\begin{equation}
\begin{array}{ccccl}
  \partial_t \Ket{\psi^\text{H}(t)} &=& -i H^\text{H}(t) \Ket{\psi^\text{H}(t)} &+& i G^\text{H}(t) \Ket{\psi^\text{H}(t)} \vspace{.15cm}\\
  \partial_t A^\text{H}(t) &=& (\partial_t A^\text{S})^\text{H}(t) &+& i [G^\text{H}(t), A^\text{H}(t)]
\end{array} \label{eq:G picture}
\end{equation}
  where $(\partial_t A^\text{S})^\text{H}(t) = U^\dagger(t) \partial_t A^\text{S}(t) U(t)$ [\eqnref{eq:Htransform}].

To obtain the Heisenberg picture, we simply choose
\begin{equation}
  G^\text{H}(t) = H^\text{H}(t) \label{eq:GH}
\end{equation}
  with the initial condition $U(0) = \mathbbl{1}$ at $t=0$,
  where $\mathbbl{1}$ denotes the identity operator.
This choice makes the wavefunction $\ket{\psi^\text{H}}$ constant in time,
  while operators evolve in time.
If the Hamiltonian is time-independent,
  then the Hamiltonian is the same in the Schr\"{o}dinger and Heisenberg pictures;
  i.e. $H^\text{S}(t) = H^\text{H}(t)$ if $\partial_t H^\text{S} = 0$.

\subsection{Locality}

A nice feature of the Heisenberg picture is that if the Hamiltonian is local,
  then the time evolution of observables is \emph{explicitly} local.
A local Hamiltonian has the form
\begin{equation}
  H = \sum_J H_J \label{eq:Hlocal}
\end{equation}
  where each $H_J$ only acts on a finite patch of space $J$.
We use capital letters, $I$, $J$, and $K$, to denote patches of space.

Now consider the time evolution of a local operator in the Heisenberg picture:
\begin{equation}
  \partial_t A_I^\text{H}(t)
    = i [H^\text{H}(t), A_I^\text{H}(t)] + (\partial_t A_I^\text{S})^\text{H}(t) \label{eq:Heisenberg}
\end{equation}
Throughout this work, $A_I$ denotes a local operator that acts within the spatial patch $I$
  (when viewed in the Schr\"{o}dinger picture),
  and similar for $B_J$, etc.
Due to locality, most Hamiltonian terms cancel out in the first term.
The result is an explicitly local Heisenberg equation of motion:
\begin{equation}
  \partial_t A_I^\text{H}(t)
    = i [H_{\langle I \rangle}^\text{H}(t), A_I^\text{H}(t)] + (\partial_t A_I^\text{S})^\text{H}(t) \label{eq:localHeisenberg}
\end{equation}
  where $(\partial_t A_I^\text{S})^\text{H}(t) = U^\dagger(t) \, (\partial_t A_I^\text{S})^\text{S}(t) \, U(t)$ [\eqnref{eq:Htransform}].
$H_{\langle I \rangle}$ is a sum over nearby Hamiltonian terms:
\begin{equation}
  H_{\langle I \rangle} = \sum_J^{J \cap I \neq \emptyset} H_J \label{eq:HI}
\end{equation}
$\sum_J^{J \cap I \neq \emptyset}$ denotes a sum over patches $J$ that have overlap with patch $I$.
Note that the local Hamiltonian terms $H_I^\text{H}(t)$ also evolve according to \eqnref{eq:localHeisenberg}.

Locality is explicit in this local Heisenberg picture because
  the time evolution of each local operator $A_I^\text{H}(t)$ only depends on nearby time-dependent operators $H_J^\text{H}(t)$.
Locality is not explicit in Schr\"{o}dinger's picture since
  the time evolution of the wavefunction globally depends on all Hamiltonian terms,
  and there is no sense in which the wavefunction (or parts of it)
  can be associated with a point in space.

\section{Gauge Picture}
\label{sec:gaugePicture}

\begin{figure}
  \centering
  \includegraphics{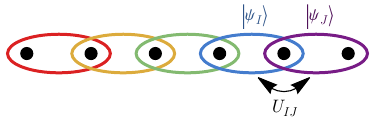}
  \caption{%
    An example of a chain of qubits (black dots) and spatial patches (colored ovals) consisting of pairs of neighboring qubits.
    A local wavefunction $\ket{\psi_I}$ is associated with each patch $I$,
      and the Hilbert spaces associated with neighboring patches are related by unitary connections $U_{IJ}$.
  }\label{fig:patches}
\end{figure}

We now want to obtain a local picture of quantum dynamics that features
  time-dependent wavefunctions and time-independent operators.
To do this, we first choose a set of local patches of space that cover the entire space (or lattice);
  see \figref{fig:patches} for an example.
Then for each patch of space,
  we apply a time-dependent unitary transformation $U_I(t)$
  to the Heisenberg picture wavefunction and local operators:
\begin{equation}
\begin{aligned}
  \bKet{\psi_I} &= U_I \bKet{\psi^\text{H}} \\
  A_I^\text{G} &= U_I A_I^\text{H} U_I^\dagger
\end{aligned} \label{eq:localGauge}
\end{equation}
We thus obtain a separate \emph{local wavefunction} $\ket{\psi_I}$ for each patch of space.
For a certain transformation $U_I$, the wavefunctions $\ket{\psi_I}$ are local in the sense that their dynamics are local and only depend on nearby Hamiltonian terms.
Note that $\ket{\psi_I}$ is a many-body wavefunction that belongs to the same Hilbert space as the usual wavefunction.
The second line transforms local operators (including $H_I$ and $H_{\langle I \rangle}$)
  from the Heisenberg picture into a new picture of quantum dynamics.
The ``G'' superscript labels time-dependent operators that evolve within this picture.
We omit the ``G'' superscript for $\ket{\psi_I}$ since
  the $\ket{\psi_I}$ notation is not used in other pictures;
  thus, there should be no confusion.
To further reduce clutter, we also suppress the ``$(t)$'' notation for time-dependent variables.

After applying this unitary transformation,
  correlation functions within a single patch become
  $\braket{\psi^\text{H} | A_I^\text{H} | \psi^\text{H}}
  =\braket{\psi_I | A_I^\text{G} | \psi_I}$.
Correlation functions of operator products acting on multiple patches becomes
\begin{equation}
\begin{aligned}
    &\Braket{\psi^\text{H} | A_I^\text{H} \quad\;\;\, B_J^\text{H} \quad\;\;\; C_K^\text{H} | \psi^\text{H}} \\
   =&\Braket{\psi_I \, | A_I^\text{G} U_{IJ} B_J^\text{G} U_{JK} C_K^\text{G} | \psi_K}
\end{aligned} \label{eq:same}
\end{equation}
  where we define
\begin{equation}
  U_{IJ} = U_I U_J^\dagger \label{eq:UIJ}
\end{equation}
A \emph{connection} $U_{IJ}$ always appears between variables associated with different spatial patches.
Equation~\eqref{eq:localGauge} implies that the following identities hold for all $t$:
\begin{equation}
\begin{aligned}
  U_{IJ} \ket{\psi_J} &= \ket{\psi_I} \\
  U_{IJ} U_{JK} &= U_{IK} 
\end{aligned} \label{eq:identities}
\end{equation}
  and $U_{IJ}^\dagger = U_{JI}$ and $U_{II} = \mathbbl{1}$.
Therefore, $U_{IJ}$ serves the role of a connection that transports the local wavefunctions between different spatial patches.
But the connections are trivial in the sense that there is no curvature or nontrivial holonomy (e.g. $U_{IJ} U_{JK} U_{KI} = \mathbbl{1}$).

Let $G_I(t)$ be the Hermitian operator such that
\begin{equation}
  \partial_t U_I = -i G_I U_I \label{eq:GU}
\end{equation}
Then the time derivatives of the local wavefunctions and operators are
\begin{equation}
\begin{aligned}
  \partial_t \ket{\psi_I} &= -i G_I \ket{\psi_I} \\
  \partial_t A_I^\text{G} &= i [H^\text{G}_{\langle I\rangle} - G_I, A_I^\text{G}] + (\partial_t A_I^\text{S})^\text{G}
\end{aligned} \label{eq:localPicture}
\end{equation}
This follows from
  $\partial_t \ket{\psi^\text{H}} = 0$
  and \eqsref{eq:localHeisenberg} and \eqref{eq:localGauge}.
$(\partial_t A_I^\text{S})^\text{G} = U_I (\partial_t A)^\text{H} U_I^\dagger$
  is defined from \eqnref{eq:localGauge}.

A nice choice for $U_I$ at $t=0$ and $G_I$ is
\begin{equation}
\begin{aligned}
  G_I &= H^\text{G}_{\langle I \rangle} \\
  U_I(0) &= \mathbbl{1}
\end{aligned}
\end{equation}
  since this makes local operators $A_I^\text{G}$ (and Hamiltonian terms)
  equal to Schr\"{o}dinger picture operators:
\begin{align}
  A_I^\text{G} &= A_I^\text{S} &
  H_I^\text{G} &= H_I^\text{S} \label{eq:sameOps}
\end{align}

We thus obtain the gauge picture equations of motion, for which
  the local wavefunctions (which are vectors in the many-body global Hilbert space) and operators evolve as follows:
\begin{equation}
\boxed{\begin{aligned}
  \partial_t \ket{\psi_I} &= -i H^\text{G}_{\langle I \rangle} \ket{\psi_I} \\
  \partial_t U_{IJ} &= - i H^\text{G}_{\langle I \rangle} U_{IJ} + i U_{IJ} H^\text{G}_{\langle J \rangle}
\end{aligned}} \label{eq:localSchrodinger}
\end{equation}
$H^\text{G}_{\langle I\rangle}$ is defined from \eqsref{eq:HI} and \eqref{eq:localGauge}
  and can be expressed as
\begin{equation}
  \boxed{H^\text{G}_{\langle I\rangle}
    = \sum_J^{J \cap I \neq \emptyset} U_{IJ} H_J^\text{G} U_{JI}} \label{eq:H'}
\end{equation}
Although the Schr\"{o}dinger and Heisenberg pictures are linear differential equations,
  the gauge picture involves \emph{non-linear} differential equations for $U_{IJ}$.
The $t=0$ initial conditions are
\begin{equation}
\begin{aligned}
  \bKet{\psi_I(0)} &= \bKet{\psi^\text{S}(0)} \\
  U_{IJ}(0) &= \mathbbl{1}
\end{aligned} \label{eq:init}
\end{equation}
Note that one does not need to keep track of a $U_{IJ}$ for every pair of $I$ and $J$;
  only considering overlapping pairs of spatial patches suffices.
If we want to describe the dynamics of a mixed state,
  then we could take $\ket{\psi^\text{S}(0)}$ to be a purification of the mixed state.

Thus, we were able to make Schr\"{o}dinger's picture explicitly local by
  introducing a local wavefunction $\ket{\psi_I}$ for each patch of space
  and unitary connections $U_{IJ}$ that map between the Hilbert spaces of the different patches.
Locality is explicit in the sense that
  the time evolution of $\ket{\psi_I}$ and $U_{IJ}$
  only depends on nearby variables ($\ket{\psi_I}$, $U_{IJ}$, and $H_J$).

In the gauge picture,
  physical correlation functions [e.g. \eqnref{eq:same}] are invariant under the following local gauge invariance:
\begin{align}
  \ket{\psi_I} &\to \Lambda_I \ket{\psi_I} & U_I &\to \Lambda_I U_I \nonumber\\
  A_I^\text{G} &\to \Lambda_I A_I^\text{G} \Lambda_I^\dagger & U_{IJ} &\to \Lambda_I U_{IJ} \Lambda_J^\dagger \label{eq:gaugeSymmetry}
\end{align}
  where $\Lambda_I$ are arbitrary unitary transformations.
For example, $\braket{\psi_I | B_J^\text{G} | \psi_I}$ and $\braket{\psi_I | A_I^\text{G} B_J^\text{G} | \psi_J}$
  are not invariant under this transformation and are therefore
  not physical in the sense that they are not equal to
  $\braket{\psi^\text{H} | B_J^\text{H} | \psi^\text{H}}$ or $\braket{\psi^\text{H} | A_I^\text{H} B_J^\text{H} | \psi^\text{H}}$
  in the Heisenberg (or Schr\"{o}dinger) picture.
Therefore, the local wavefunctions $\ket{\psi_I}$ are local in the sense that they can only be used to
  calculate local expectation values of local operators $A_I$ acting within the associated spatial patch,
  unless we make use of connections $U_{IJ}$ to connect different patches.
The above gauge transformation can be used to transform the gauge picture to the Heisenberg picture (with $U_{IJ} = \mathbbl{1}$) by taking $\Lambda_I = U_I^\dagger$
  or to the Schr\"odinger picture by taking $\Lambda_I = U U_I^\dagger$.

Note that \eqnref{eq:gaugeSymmetry} generalizes the usual global unitary invariance in quantum mechanics (where $\ket{\psi} \to U \ket{\psi}$ and $A \to U A U^\dagger$)
  to a local unitary invariance.
Due to the appearance of this local gauge invariance,
  we call this picture of quantum dynamics the \emph{gauge picture}.

In order to extract the Schr\"odinger picture wavefunction from the gauge picture,
  one can use the correlation functions to first calculate the density matrix.
For example, for a system of $n$ qubits, the density matrix is
\begin{align}
  \rho = 2^{-n} \sum_{\mu_1 \cdots \mu_n} &
    \sigma_1^{\mu_1} \cdots \sigma_n^{\mu_n} \label{eq:rho}\\
    & \;\; \braket{\Psi_1 | \, \sigma_1^{\mu_1} \, U_{1,2} \, \sigma_2^{\mu_2} \cdots \, U_{n-1,n} \, \sigma_n^{\mu_n} | \Psi_n} \nonumber
\end{align}
  where $\sigma_i^{\mu}$ are Pauli operators.
Here we take each patch to consist of just a single qubit
  so that $I=1,\ldots,n$ indexes the qubits/patches along some path.
If the gauge picture is initialized to a pure state using \eqnref{eq:init},
  then $\rho$ will always be pure,
  and the wavefunction can be uniquely obtained from $\rho$ (up to a phase)
  as the only eigenvector of $\rho$ with nonzero eigenvalue.
More generally however, the gauge picture can encode a mixed state
  [as explained below \eqnref{eq:init}].

As an alternative to \eqnref{eq:localSchrodinger},
  local dynamics can also be calculated by integrating
\begin{equation}
  \partial_t U_I = - i H_{\langle I \rangle}^\text{G} U_I \label{eq:localSchrodingerUI}
\end{equation}
  [from \eqnref{eq:GU}] with initial condition $U_I(0) = \mathbbl{1}$.
Equation~\eqref{eq:localSchrodingerUI} can be rewritten more explicitly as
\begin{equation}
  \partial_t U_I = - i \sum_J^{J \cap I \neq \emptyset} U_I (U_J^\dagger H_J^\text{G} U_J) \label{eq:dUI}
\end{equation}
Since wavefunctions are constant in the Heisenberg picture,
  \eqnref{eq:localGauge} implies that the time-evolved local wavefunction is
\begin{equation}
  \bKet{\psi_I(t)} = U_I(t) \bKet{\psi^\text{S}(0)} \label{eq:UPsi}
\end{equation}

\subsection{Solving the Equations of Motion}

Let us derive an expression for $U_I$ in terms of just the Hamiltonian.
Simplifying \eqnref{eq:dUI} using
  $H_J^\text{G} = U_J H_J^\text{H} U_J^\dagger$ [\eqnref{eq:localGauge}] and
  $H_{\langle I \rangle} = \sum_J^{J \cap I \neq \emptyset} H_J$ [\eqnref{eq:HI}] yields
\begin{align}
  \partial_t U_I &= - i U_I H_{\langle I \rangle}^\text{H} \nonumber\\
   &= - i \widetilde{U}_I^\dagger H_{\langle I \rangle}^\text{S} U \label{eq:dUI2}
\end{align}
The second line follows from
  $H_{\langle I \rangle}^\text{H} = U^\dagger H_{\langle I \rangle}^\text{S} U$
  in \eqnref{eq:Htransform} after we define
\begin{equation}
  \widetilde{U}_I = U U_I^\dagger
\end{equation}
The derivative of $\widetilde{U}_I$ is
\begin{equation}
  \partial_t \widetilde{U}_I = -i \left(H^\text{S} - H_{\langle I \rangle}^\text{S}\right) \widetilde{U}_I \label{eq:dtU}
\end{equation}
  where $H^\text{S} - H_{\langle I \rangle}^\text{S}$ is the sum of Hamiltonian terms that do not overlap with patch $I$.
Equation~\eqref{eq:dtU} follows from \eqnref{eq:dUI2} and
\begin{equation}
  \partial_t U(t) = -i H^\text{S}(t) U(t) \label{eq:dU}
\end{equation}
  [from \eqsref{eq:dUG} and \eqref{eq:GH}].

We now have linear differential equations for
  $U(t)$ and $\widetilde{U}_I(t)$ [\eqsref{eq:dU} and \eqref{eq:dtU}].
For time-independent Hamiltonians, we can solve these differential equations:
\begin{equation}
\begin{aligned}
  U(t) &= e^{-iHt} \\
  \widetilde{U}_I(t) &= e^{-i(H - H_{\langle I \rangle}^\text{S})t}
\end{aligned} \label{eq:tU}
\end{equation}
Once we have calculated $U$ and $\widetilde{U}_I$,
  we can obtain $U_I$ from:
\begin{equation}
  U_I = \widetilde{U}_I^\dagger U \label{eq:UI}
\end{equation}
  which follows from the definition of $\widetilde{U}_I = U U_I^\dagger$.
Equation~\eqref{eq:UI} is useful when
  we want to calculate $U_I$, $\ket{\psi_I}$ [from \eqnref{eq:localGauge}], or $U_{IJ}$ [from \eqnref{eq:UIJ}]
  without caring about explicitly local equations of motion.

$U(t)$ describes the time evolution of the Schr\"{o}dinger picture wavefunction:
  $\ket{\psi^\text{S}(t)} = U(t) \ket{\psi^\text{S}(0)}$.
Therefore, \eqnref{eq:UI} tells us that $\widetilde{U}_I$ describes how much the gauge picture has deviated from the Schr\"{o}dinger picture.
Indeed, the wavefunctions of the two pictures are related by $\widetilde{U}_I$:
\begin{equation}
  \ket{\psi_I} = \widetilde{U}_I^\dagger \ket{\psi^\text{S}} \label{eq:psiGS}
\end{equation}
This identity follows from comparing \eqnref{eq:UPsi} to $\ket{\psi^\text{S}(t)} = U(t) \ket{\psi^\text{S}(0)}$.
The role of $\widetilde{U}_I$ is to cancel out the effects of distant Hamiltonian terms so that $U_I$ evolves locally.
If we apply the gauge transformation \eqref{eq:gaugeSymmetry} with $\Lambda_I = \widetilde{U}_I$,
  then all gauge picture variables transform back into the Schr\"{o}dinger picture:
  $\ket{\psi_I} \to \ket{\psi^\text{S}}$,
  $U_I \to U$, and
  $U_{IJ} \to \mathbbl{1}$.
Local operators $A_I^\text{G} = A_I^\text{S}$ are already equal in the two pictures,
  and they are invariant since they commute with $\widetilde{U}_I$.

\subsection{Generalized Hamiltonians}

It is also possible to handle Hamiltonian terms that are not supported on a single spatial patch.
This is useful when we want the spatial patches to consist of just single lattice site.

Consider a Hamiltonian
\begin{equation}
  H = \sum_{J\cdots K} \sum_{\mu\cdots\nu}
      h_{J\cdots K}^{\mu\cdots\nu} \; \tau^\mu_J \cdots \tau^\nu_K \label{eq:Hgen}
\end{equation}
  that consists of a sum $\sum_{J\cdots K}$ over multiple spatial patches $J \cdots K$.
The $h_{J\cdots K}^{\mu\cdots\nu}$ are (generically time-dependent) real coefficients,
  and $\tau^\mu_J$ denotes an operator (in the Schr\"{o}dinger or gauge picture)
  indexed by $\mu$ with support on patch $J$.
For example, we could take the patches to consist of a single qubit,
  and the $\tau^\mu_J$ could be Pauli operators.
For local Hamiltonians, $h_{J\cdots K}^{\mu\cdots\nu}$ will only be nonzero if the patches $J\cdots K$ are close together.

Equation~\eqref{eq:H'} then generalizes to
\begin{align}
  H_{\langle I\rangle}^\text{G}
    = \sum_{J\cdots K}^{(J\cup\cdots K)\cap I \neq \emptyset} \sum_{\mu\cdots\nu} \,&h_{J\cdots K}^{\mu\cdots\nu}
    \label{eq:H'gen} \\ &
          (U_{IJ} \tau^\mu_J U_{JI}) \cdots (U_{IK} \tau^\nu_K U_{KI}) \nonumber
\end{align}
  where $\sum_{J\cdots K}^{(J\cup\cdots K)\cap I \neq \emptyset}$ sums over spatial patches $J\cdots K$
  such that the union $J\cup\cdots K$ has nontrivial overlap with patch $I$.
The gauge picture equations of motion [\eqnref{eq:localSchrodinger} or \eqref{eq:localSchrodingerUI}]
  can then be applied using \eqnref{eq:H'gen}
  [instead of \eqnref{eq:H'}].
This can be shown by rederiving the local Heisenberg and gauge pictures for
  the Hamiltonian in \eqnref{eq:Hgen} [instead of \eqnref{eq:Hlocal}].

\section{Local Interaction Picture}
\label{sec:interactionPicture}

In this section, we equate variables in the gauge picture
  to variables in the interaction picture with interacting Hamiltonian $H_{\langle J \rangle}$ [\eqnref{eq:HI}],
  which is the sum over Hamiltonian terms with some support on patch $J$.
In particular, we show that the gauge picture local wavefunction $\ket{\psi_J}$ is equal to this interaction picture's wavefunction.

In the interaction picture, the Hamiltonian
\begin{equation} H = H_0 + H_1 \label{eq:H interaction} \end{equation}
  is divided into
  interacting $H_1$ and non-interacting $H_0$ parts.
The interaction picture wavefunction and operators are defined as
\begin{equation}
\begin{aligned}
  \Ket{\psi^\text{I}} &= U_0^\dagger \Ket{\psi^\text{S}} \\
  A^\text{I} &= U_0^\dagger A^\text{S} U_0
\end{aligned} \label{eq:interactionPsi}
\end{equation}
  where $U_0(t)$ is the solution to
\begin{equation}
  \partial_t U_0 = -i H_0^\text{S} U_0 \label{eq:dU0}
\end{equation}
  with $t=0$ initial condition $U_0(0) = \mathbbl{1}$.
The utility of the interaction picture is that wavefunctions evolve via the interacting Hamiltonian
  while operators evolve via the non-interacting Hamiltonian:
\begin{equation}
\begin{aligned}
  \partial_t \Ket{\psi^\text{I}} &= -i H_1^\text{I} \Ket{\psi^\text{I}} \\
  \partial_t A^\text{I} &= +i [H_0^\text{I}, A^\text{I}] + (\partial_t A^\text{S})^\text{I}
\end{aligned} \label{eq:interactionPicture}
\end{equation}
  where $H_0^\text{I} = U_0^\dagger H_0^\text{S} U_0$ and $H_1^\text{I} = U_1^\dagger H_1^\text{S} U_1$
  also obey the second line,
  and $(\partial_t A^\text{S})^\text{I} = U_0^\dagger \partial_t A^\text{S} U_0$ [as defined by \eqnref{eq:interactionPsi}].
If $H_0^\text{S}$ is time-independent, then $H_0^\text{I} = H_0^\text{S}$.

Consider a fixed spatial patch $J$, and
  let the interaction Hamiltonian be
\begin{equation} H_1 = H_{\langle J \rangle} \label{eq:HI1} \end{equation}
Then $\widetilde{U}_J$ and $U_0$ are equal since they obey the same equations of motion,
  $\partial_t \widetilde{U}_J = -i (H^\text{S}-H_{\langle J \rangle}^\text{S}) \widetilde{U}_J = -i H_0^\text{S} \widetilde{U}_J$
  [by \eqsref{eq:dtU}, \eqref{eq:H interaction}, and \eqref{eq:HI1}]
  and $\partial_t U_0 = -i H_0^\text{S} U_0$ [\eqnref{eq:dU0}], with the same initial conditions
  $\widetilde{U}_J(0) = U_0(0) = \mathbbl{1}$.
Inserting $\widetilde{U}_J = U_0$ into $U_J = \widetilde{U}_J^\dagger U$ [\eqnref{eq:UI}] yields
\begin{equation} U_J = U_0^\dagger U \label{eq:interactionUJ} \end{equation}
We thus arrive at an equivalence between the gauge picture on patch $J$
  and this interaction picture:
\begin{equation}
\begin{aligned}
  \bKet{\psi_J} &= \bKet{\psi^\text{I}} \\
  A_J^\text{G} &= A_J^\text{S} = A_J^\text{I}
\end{aligned}
\end{equation}
The first line follows from
  $\ket{\psi_J} = U_J \ket{\psi^\text{H}} = U_J U^\dagger \ket{\psi^\text{S}} = U_0^\dagger \ket{\psi^\text{S}} = \ket{\psi^\text{I}}$ [using \eqsref{eq:localGauge}, \eqref{eq:Htransform}, \eqref{eq:interactionUJ}, and \eqref{eq:interactionPsi}].
The equalities in the second line respectively follow from
  \eqnref{eq:sameOps} and $A_J^\text{I} = U_0^\dagger A_J^\text{S} U_0$ [\eqnref{eq:interactionPsi}]
  since $U_0$ and $A_J^\text{S}$ commute when $H_1 = H_{\langle J \rangle}$.

The above relation makes sense:
$\ket{\psi_J}$ and $\ket{\psi^\text{I}}$
  respectively evolve via $H_{\langle J \rangle}^\text{G}$ and $H_1^\text{I}$;
  therefore $\ket{\psi_J} = \ket{\psi^\text{I}}$ since $H_{\langle J \rangle} = H_1$.
Operators in the interaction picture evolve via the non-interacting Hamiltonian $H_0^\text{I}$,
  which commutes with operators supported on $J$;
  therefore operators $A_J$ supported on $J$ are equal in the Schr\"{o}dinger and interaction pictures.

Unlike the gauge picture,
  the above interaction picture is not explicitly local
  since $H_1^\text{1}$ evolves via a global Hamiltonian $H_0^\text{I}$.
However, the gauge picture could be viewed as many simultaneous interactions pictures,
  one for each spatial patch $J$ and interaction Hamiltonian $H_1 = H_{\langle J \rangle}$,
  but with modified equations of motion to obtain explicitly local dynamics.

\section{Quantum Circuits}
\label{sec:circuits}

\begin{figure*}
  \centering
  \subfloat[Schr\"{o}dinger: \\ $\ket{\psi^\text{S}(t_\text{f})} = U \ket{\psi_0}$]{\includegraphics{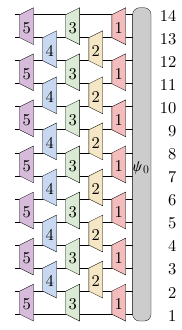}} \hspace{1cm}
  \subfloat[Heisenberg: \\ $A^\text{H}(t_\text{f}) = U A(0) \, U^\dagger$]{\includegraphics{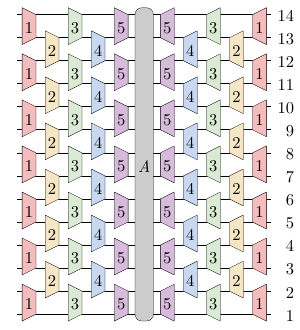}} \hspace{1cm}
  \subfloat[local Heisenberg: \\ $A_I^\text{H}(t_\text{f}) = U A_I(0) U^\dagger$ \label{fig:localHeisenberg}]{\includegraphics{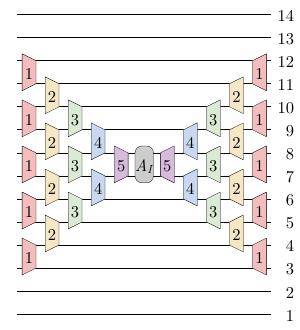}}
  \caption{%
    Dynamical variables used in the standard pictures of quantum dynamics after the time evolution of a quantum circuit.
    (a) The Schr\"{o}dinger picture wavefunction $\ket{\psi^\text{S}(t_\text{f})} = U \ket{\psi_0}$
      after the quantum circuit $U=U(t_\text{f})$ is applied to the initial wavefunction $\ket{\psi_0}$.
    Each trapezoid depicts a unitary operator acting on two qubits from time $t=\tau-1$ to $\tau$,
      where $\tau$ is the number inside the trapezoid.
    The circuit $U$ acts until $t_\text{f}=5$, and consists of the composition of these unitaries.
    Qubits are numbered on the right.
    (b) A time-evolved operator $A^\text{H}(t_\text{f}) = U A(0) \, U^\dagger$ in the Heisenberg picture.
    Trapezoids reflected horizontally denote the inverses of the associated 2-qubit unitaries.
    (c) For a local operator $A_I^\text{H}(t)$,
      many of the unitaries cancel with their inverse.
    Here, $A_I(0)$ acts on a spatial patch $I=(7,8)$ consisting of two qubits.
  }\label{fig:circuits}
\end{figure*}

\begin{figure*}
  \centering
  \subfloat[$U_I(t_\text{f})$ \label{fig:UI}]{\includegraphics{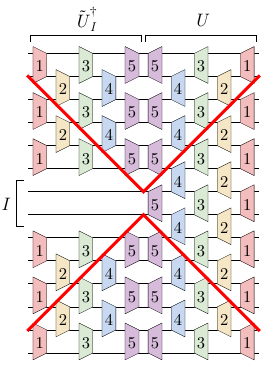}
                                              \hspace{.1cm} \raisebox{2.65cm}{\scalebox{2}{=}} \hspace{.1cm}
                                              \includegraphics{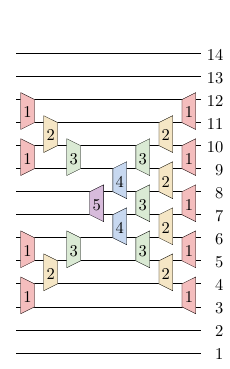}} \hspace{2.6cm}
  \subfloat[$U_J^\dagger(t_\text{f})$ \label{fig:UJ}]{\includegraphics{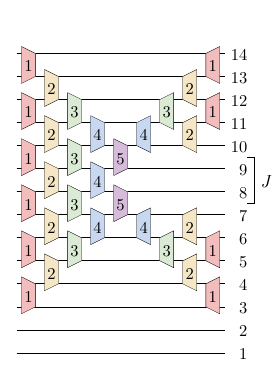}} \\
  \subfloat[$U_{IJ}(t_\text{f})$ \label{fig:UIJ}]{\includegraphics{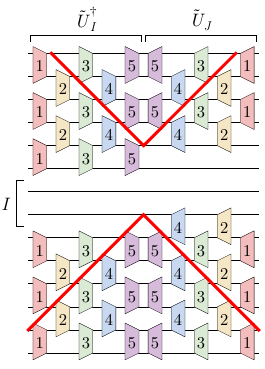}
                                                  \hspace{.1cm} \raisebox{2.65cm}{\scalebox{2}{=}} \hspace{.1cm}
                                                  \includegraphics{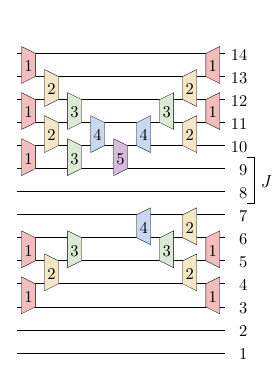}}
  \caption{%
    Dynamical variables in the gauge picture after the same quantum circuit as in \figref{fig:circuits}.
    (a) The unitary $U_I(t_\text{f})$ with $I=(7,8)$.
      We also pictorially show that $\widetilde{U}_I^\dagger U = U_I$ [\eqnref{eq:UI}]
      after many unitaries outside the red light cone cancel with their inverse.
    $\widetilde{U}_I$ is the same as $U$,
      but with all unitaries (or Hamiltonian terms) that overlap with the spatial patch $I$ removed.
    The local wavefunction follows from $\ket{\psi_I} = U_I \ket{\psi_0}$.
    (b) The unitary $U_J^\dagger(t_\text{f})$ with $J=(8,9)$.
    (c) The connection $U_{IJ}(t_\text{f}) = U_I(t_\text{f}) U_J^\dagger(t_\text{f})$,
      which is also equal to $\widetilde{U}_I^\dagger \widetilde{U}_J$.
  }\label{fig:gaugeCircuits}
\end{figure*}

In this section, we study how unitary operators act in the gauge picture.
As an instructive example,
  in \figsref{fig:circuits} and \ref{fig:gaugeCircuits} we depict dynamical variables used in different pictures of quantum dynamics
  after the time evolution of a quantum circuit.
The quantum circuit consists of a composition of unitary operators that each act on pairs of neighboring qubits.
Equivalently, the circuit could result from a time-dependent Hamiltonian evolution.

Similar to operators in the local Heisenberg picture (\figref{fig:localHeisenberg}),
  the unitaries $U_I$ and $U_{IJ}$ (\figref{fig:UI} and \ref{fig:UIJ}) in the gauge picture also grow as time evolves.
\figsref{fig:localHeisenberg} and \ref{fig:gaugeCircuits} demonstrate the explicit locality of the local variables in the Heisenberg and gauge pictures:
  over a short time evolution,
  these local variables are not affected by anything far away.
\figref{fig:UI} demonstrates that this locality is achieved via the composition $\widetilde{U}_I^\dagger U$,
  where $\widetilde{U}_I^\dagger$ cancels out distant unitary evolution.

\subsection{Unitary Operators}

Let us describe the affect of local unitary operators in the gauge picture using equations.
Consider the action of many local unitary operators $u_I$ on the Schr\"{o}dinger picture wavefunction:
\begin{equation}
  \ket{\psi^\text{S}} \to \prod_I u_I \ket{\psi^\text{S}}
\end{equation}
Assume that each unitary $u_I$ acts within patch $I$ and that all unitaries commute
  (e.g. as they do for each layer of unitaries in \figref{fig:circuits}):
\begin{equation}
  [u_I,u_J] = 0
\end{equation}

The local gauge picture variables transform as follows:
\begin{equation}
\begin{aligned}
  \ket{\psi_I} &\to u_{\langle I \rangle} \ket{\psi_I} \\
  U_{IJ} &\to u_{\langle I \rangle} U_{IJ} u_{\langle J \rangle}^\dagger \\
  U_I &\to u_{\langle I \rangle} U_I
\end{aligned} \label{eq:unitary}
\end{equation}
In analogy to $H_{\langle I \rangle}^\text{G}$, we define
\begin{equation}
  u_{\langle I \rangle} = \prod_J^{J \cap I \neq \emptyset} U_{IJ} u_J U_{JI} \label{eq:uI'}
\end{equation}
  where $\prod_J^{J \cap I \neq \emptyset}$ denotes a product over all patches that overlap with patch $I$.
Note that $U_{IJ} = U_I U_J^\dagger = \widetilde{U}_I^\dagger \widetilde{U}_J$
  since $U_I = \widetilde{U}_I^\dagger U$ [\eqnref{eq:UI}], which implies that
  $U_{IJ} u_J U_{JI} = \widetilde{U}_I^\dagger u_J \widetilde{U}_I$
  since $u_J$ and $\widetilde{U}_J$ commute.
Therefore, all of the $U_{IJ} u_J U_{JI}$ terms in \eqnref{eq:uI'} commute, and
  $u_{\langle I \rangle}$ can also be expressed as
\begin{equation}
  u_{\langle I \rangle} = \widetilde{U}_I^\dagger \left( \prod_J^{J \cap I \neq \emptyset} u_J \right) \widetilde{U}_I
\end{equation}

To derive \eqnref{eq:unitary},
  imagine that the local unitaries $u_I = e^{-iH_It_0}$ are applied using a time-independent Hamiltonian
  $H = \sum_I H_I$ with commuting local terms (i.e. $[H_I,H_J]=0$).
Then we can use \eqsref{eq:dU} and \eqref{eq:dtU} to calculate how $U$ and $\widetilde{U}_I$ transform:
\begin{equation}
\begin{aligned}
  U &\to \big(\textstyle \prod_J u_J\big) \, U \\
  \widetilde{U}_I &\to \left(\prod_J^{J \cap I = \emptyset} u_J\right) \widetilde{U}_I
\end{aligned}
\end{equation}
  where $\prod_J^{J \cap I = \emptyset}$ denotes a product over all patches that do not overlap with patch $I$.
This allows us to calculate the transformation
\begin{equation}
  U_I \to \widetilde{U}_I^\dagger \left(\prod_J^{J \cap I \neq \emptyset} u_J\right) U = u_{\langle I \rangle} U_I
\end{equation}
  from $U_I = \widetilde{U}_I^\dagger U$ [\eqnref{eq:UI}].
The other transformations in \eqnref{eq:unitary} are extracted from
  $\ket{\psi_I(t)} = U_I(t) \ket{\psi^\text{S}(0)}$ [\eqnref{eq:UPsi}] and $U_{IJ} = U_I U_J^\dagger$ [\eqnref{eq:UIJ}].

\section{Measurements}
\label{sec:measurements}

In the Schr\"{o}dinger picture,
  a measurement results in a nonlocal collapse of the wavefunction.
Most generally, a measurement process can be specified in terms of a collection of Kraus operators $E^{(k)}$,
  whose squares sum to the identity:
\begin{equation}
  \sum_k E^{(k)\dagger} E^{(k)} = \mathbbl{1} \label{eq:KrausSum}
\end{equation}
For example, we could take the Kraus operators to be projection operators $E^{(k)} = \ket{k} \bra{k}$
  that project onto the different eigenstates of a Hermitian operator $\sum_k \lambda_k \ket{k} \bra{k}$.
After the measurement, the wavefunction changes to
\begin{equation}
  \Ket{\psi^\text{S}} \to P_k^{-1/2} E^{(k)} \Ket{\psi^\text{S}}
\end{equation}
  with probability
\begin{equation}
  P_k = \braket{\psi^\text{S}|E^{(k)\dagger} E^{(k)}|\psi^\text{S}}
\end{equation}
Note that if all measurement outcomes are averaged over,
  the expectation value of an operator after a measurement is not affected due to \eqnref{eq:KrausSum}.
  % $\sum_k P_k \braket{\psi_k^\text{S}|\psi_k^\text{S}} = \braket{\psi^\text{S}|\psi^\text{S}}$.

Let us now consider the result of a local measurement in the gauge picture.
We specify the local measurement on a spatial patch $I_0$ using a collection of local Kraus operators $E_{I_0}^{(k)}$
  that act on the patch $I_0$.
Similar to above, the squared Kraus operators must sum to the identity:
\begin{equation}
  \sum_k E_{I_0}^{(k)\dagger} E_{I_0}^{(k)} = \mathbbl{1}
\end{equation}
After the measurement, the local wavefunction at patch $I_0$ changes to
\begin{equation}
  \bKet{\psi_{I_0}} \to \bKet{\psi^{(k)}_{I_0}} = P_k^{-1/2} E_{I_0}^{(k)} \bKet{\psi_{I_0}}
\end{equation}
  with probability
\begin{equation}
  P_k = \bBraket{\psi_{I_0}\big|E_{I_0}^{(k)\dagger} E_{I_0}^{(k)}\big|\psi_{I_0}}
\end{equation}
The connections $U_{IJ}$ are not affected.
In order to maintain consistency with the other spatial patches,
  all other local wavefunctions must transform as well:
\begin{equation}
  \bKet{\psi_J} \to U_{JI_0} \bKet{\psi^{(k)}_{I_0}} \label{eq:gaugeCollapse}
\end{equation}
Equation~\eqref{eq:psiGS} can be used to show that these equations are consistent with the Schr\"{o}dinger picture.
Equation~\eqref{eq:gaugeCollapse} demonstrates that the wavefunction collapse is nonlocal in the gauge picture.

\section{Generalized Locality}
\label{sec:patches}

In this section, we generalize the notion of locality in the gauge picture.
The underlying role of a spatial patch is to define a subspace of operators that only act within the spatial patch.
Thus, we can generalize the notion of spatial patches to subspaces of operators.
That is, instead of letting capital $I$, $J$, and $K$ denote spatial patches,
  we could instead let them denote operator subspaces.
The above equations that applied for local operators $A_I$ are generalized to apply to any operator in the subspace $I$.
We can therefore generalize the notion of geometric locality to locality between different subspaces of operators,
  which might e.g. correspond to different energy or charge densities.

We say that two operator subspaces ($I$ and $J$) do not commute
  if there exists a pair of operators ($A$ and $B$),
  with one from each subspace ($A\in I$ and $B\in J$),
  such that the pair does not commute ($[A,B] \neq 0$).
The notion of overlapping spatial patches generalizes to non-commuting operator subspaces.
Therefore in \eqnref{eq:H'},
  the sum $\sum_J^{J \cap I \neq \emptyset}$ generalizes to a sum over all operator subspaces $J$
  that do not commute with subspace $I$.
Similarly, the sum $\sum_{J\cdots K}^{(J\cup\cdots K)\cap I \neq \emptyset}$ in \eqnref{eq:H'gen}
  generalizes to a sum over all subspaces $J\cdots K$
  where any one of these subspaces does not commute with subspace $I$.

Despite the abstractness of this generalization,
  it does have physical applications.
For example, bosonic Hamiltonians (e.g. the Bose-Hubbard model)
  consisting of bosonic creation and annihilation operators
  typically exhibit a sense of locality in boson number.
That is, for many bosonic Hamiltonians, the action of a Hamiltonian term typically only changes
  the boson numbers at each site by a small amount.
Therefore, we could consider choosing the operator subspaces $I_{i,n}$ to consist of the subspace of operators
  that act at site $i$ and only on states with boson number $n$ or $n+1$ at site $i$.
To be concrete, let $\ket{n'}_i \bra{n}_i$ denote a boson operator that projects a
  wavefunction onto the subspace with $n$ bosons at site $i$
  and then changes the boson number at site $i$ to $n'$.
With this notation, the boson number operator at site $i$ could be expressed as $\hat{n}_i = \sum_{n=0}^\infty n \ket{n}_i \bra{n}_i$
  and the annihilation operator as $\hat{b}_i = \sum_{n=1}^\infty \sqrt{n} \ket{n-1}_i \bra{n}_i.$
Then in this example, the subspace $I_{i,n}$ consists of the operators spanned by
  $\ket{n}_i\bra{n}_i$, $\ket{n}_i\bra{n+1}_i$, $\ket{n+1}_i\bra{n}_i$, and $\ket{n+1}_i\bra{n+1}_i$.
In the resulting gauge picture, the operator subspaces $I_{i,n}$ and
  local wavefunctions $\ket{\psi_{I_{i,n}}}$
  are indexed by position $i$ and local boson number $n$,
  which results in equations of motion that are explicitly local in both space and local boson number.

We could similarly use the total energy or local energy density to specify the operator subspaces.
More generally, we can obtain an extra dimension of locality whenever a model has a conserved (or approximately conserved) charge or energy.
For example, consider a collection of spatial patches and a local Hamiltonian $H=\sum_I H_I$
  where each Hamiltonian term $H_I$ has many (possibly infinitely many) energy levels.
Then we could define operator subspaces $\widetilde{I}_{I,E}$ that consist of the subspace of operators that act within the spatial patch $I$
  and only on eigenstates of $H_I$ with energy between $E \pm \Delta E$
  (with $\Delta E$ chosen to be sufficiently large that we can capture the necessary operators).
In this example, the gauge picture exhibits explicit locality in both space and local energy density.

\section{Conclusion}

Having different pictures for understanding the same physics has frequently proved to be invaluable.
Furthermore, it is useful when a fundamental property of the physics is explicit in the equations,
  e.g. as in a Lorentz-invariant Lagrangian.
To this end, we studied a picture of quantum dynamics that makes locality explicit in the
  equations of motions for local wavefunctions.
From another point of view, the gauge picture can be thought of as the result of ``gauging'' the global unitary invariance of Schr\"{o}dinger's equation into a local unitary invariance.
The gauge picture open several intriguing directions for future research.

\subsection{New Approximation Technique}

Simulating the gauge picture dynamics on a computer can be achieved using a computational time that scales as $O(n_\text{U} N^3 T_\text{f} / \delta_\text{t})$,
  where $n_\text{U}$ is the number of connections
  (which typically scales with the system size),
  $N$ is the Hilbert space dimension,
  $N^3$ is the time complexity for $N\times N$ matrix multiplication in practice\footnote{%
    For extremely large $N$, an $N^\omega$ time complexity for matrix multiplication is in principle possible for some $2 \leq \omega < 3$ \cite{quantaMatrix,Strassen1969}.},
  $T_\text{f}$ is the final time,
  and $\delta_\text{t}$ is the time step used for numerical integration.\footnote{%
    The time complexity is dominated by the time it takes to multiply a connection $U_{IJ}$ with another $N\times N$ matrix in \eqsref{eq:localSchrodinger} and \eqref{eq:H'},
      which must be done $O(n_\text{U} T_\text{f} / \delta_\text{t})$ times.}
Although this is more expensive than simulating the Schr\"odinger picture,
  in another work \cite{gaugeNetworks}
  we show that the gauge picture
  facilitates a new kind of approximation scheme for improved computational efficiency.

In the Schr\"odinger picture,
  truncating the Hilbert space to low-energy states is a common technique to
  decrease the Hilbert space dimension.
For example, tight binding models
  truncate the Hilbert space to only include the energy bands closest to the chemical potential.
However, this truncation technique still requires a Hilbert space dimension that scales exponentially with system size.

\begin{figure}
  \centering
  \subfloat[matrix product state]{\includegraphics{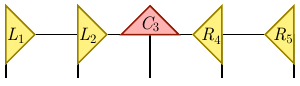}} \\
  \subfloat[Hilbert space truncation operator]{\includegraphics{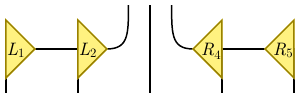}}
  \caption{%
    (a) A matrix product state in canonical form, where the yellow tensors are isometry tensors.
    The red tensor can be viewed as a local wavefunction in a truncated Hilbert space.
    (b) Removing the red tensor from the matrix product state yields the operator that maps the full Hilbert space (bottom) to the truncated Hilbert space (top).
    For more details, see Appendix B of \refcite{gaugeNetworks} (from which this figure was adapted).
  }\label{fig:MPS}
\end{figure}

The gauge picture allows a more efficient Hilbert space truncation where a different subspace of states is kept for different patches of space.
The utility of a position-dependent Hilbert space truncation is exemplified by
  matrix product states \cite{OrusTNlong} (and isometric tensor networks \cite{IsometricTN}),
  which are efficient tensor network representations of wavefunctions with area-law entanglement \cite{OrusTNlong}.
Figure~\ref{fig:MPS} (adapted from \refcite{gaugeNetworks})
  shows how a matrix product state can be viewed as a wavefunction in a position-dependent truncated Hilbert space.
Although position-dependent truncated Hilbert spaces are already used by matrix product states and isometric tensor networks,
  the gauge picture provides a new way to approximately time evolve wavefunctions using position-dependent truncated Hilbert spaces.
Compared to tensor network methods, the gauge picture provides significantly simpler quantum dynamics algorithms for models in two or more spatial dimensions and for models with fermions. 
See \refcite{gaugeNetworks} for a more thorough exploration of this approach.

\subsection{New Deformation of Quantum Mechanics}

The gauge picture agrees with the Schr\"{o}dinger picture when there is no curvature in the gauge connections, i.e. when \eqnref{eq:identities} is satisfied.
Adding curvature to the initial conditions [i.e. adding violations of \eqnref{eq:identities}]
  yields a new kind of deformation of quantum theory,
  which could yield testable experimental predictions for new physics beyond quantum theory.
Considering gauge curvature deformations of quantum theory may be appealing
  since gauge curvature has been a major theme in the Standard Model of particle physics.
Furthermore, this deformation explicitly maintains locality, in contrast to Weinberg's nonlinear deformations of quantum theory \cite{WeinbergNonlinear} which allows supraluminal communications \cite{GisinNonlinearEPR, PolchinskiNonlinearEPR}.
However, future research is needed to determine if testable predictions \cite{SlagleTestingQuantum} can be made that are consistent with previous experiments.

\subsection{Other Future Directions}

Other intriguing future directions include:
(1)~It is not clear how to describe quantum channels or the Lindblad master equation in the gauge picture.
The obstruction is that we used a unitary operator ($\widetilde{U}_I$)
  to cancel out distant unitary dynamics in order to achieve locality;
  but unitary operators can not cancel out the non-unitary dynamics of quantum channels or the Lindbladian.
(2)~Does the gauge picture have useful applications for
  understanding exotic dynamical many-body phenomena, such as
  out-of-time-ordered correlation functions \cite{SwingleOTOC,OTOCRev},
  many body localization \cite{MBL1,MBL2},
  or the speed of information propagation \cite{LiebRobinsonRev,LiebRobinsonButterfly,TighteningLiebRobinson}?

\begin{acknowledgments}
We thank Lesik Motrunich, Sayak Guha Roy, Sagar Vijay, Gunhee Park, and Garnet Chan for helpful conversations.
K.S. was supported by
  the Walter Burke Institute for Theoretical Physics at Caltech; and
  the U.S. Department of Energy, Office of Science, National Quantum Information Science Research Centers, Quantum Science Center.
This research was supported in part by the National Science Foundation Grant No. NSF PHY-1748958 and the Gordon and Betty Moore Foundation Grant No. 2919.02. % KITP Quantum Crystal program
We also acknowledge funding from the Welch Foundation through Grant No. C-2166-20230405.
\end{acknowledgments}

\bibliographystyle{quantum}
\bibliography{gaugePicture}

\end{document}